# Improved Child Text-to-Speech Synthesis through Fastpitch-based Transfer Learning

Rishabh Jain
*C3 Imaging Research Center*
*University of Galway*
Galway, Ireland
rishabh.jain@universityofgalway.ie

Peter Corcoran
*C3 Imaging Research Center*
*University of Galway*
Galway, Ireland
peter.corcoran@universityofgalway.ie

*Abstract*— Speech synthesis technology has witnessed significant advancements in recent years, enabling the creation of natural and expressive synthetic speech. One area of particular interest is the generation of synthetic child speech, which presents unique challenges due to children's distinct vocal characteristics and developmental stages. This paper presents a novel approach that leverages the Fastpitch text-to-speech (TTS) model for generating high-quality synthetic child speech. This study uses the transfer learning training pipeline. The approach involved finetuning a multi-speaker TTS model to work with child speech. We use the 'cleaned' version of the publicly available MyST dataset (55 hours) for our finetuning experiments. We also release a prototype dataset of synthetic speech samples generated from this research together with model code to support further research. By using a pretrained MOSNet, we conducted an objective assessment that showed a significant correlation between real and synthetic child voices. Additionally, to validate the intelligibility of the generated speech, we employed an automatic speech recognition (ASR) model to compare the word error rates (WER) of real and synthetic child voices. The speaker similarity between the real and generated speech is also measured using a pretrained speaker encoder.

I. INTRODUCTION

Speech synthesis technology has witnessed significant advancements in recent years, enabling the creation of natural and expressive synthetic speech. One area of particular interest is the generation of synthetic child speech, which presents unique challenges due to children's distinct vocal characteristics and developmental stages. Early research on Text-to-Speech (TTS) synthesis began several decades ago, primarily using concatenative and parametric methods [1]–[4]. While these methods generated speech from text, the resulting audio lacked naturalness and sounded robotic. Recent advancements in TTS models, mainly based on deep neural networks (DNN), have significantly improved the quality of synthesized speech. Tacotron [5], a neural sequence-to-sequence model, marked a notable improvement in speech synthesis quality. Subsequent models like Tacotron2 [6], FastSpeech [7], FastSpeech2 [8], FlowTTS [9], GlowTTS [10], Fastpitch [11], and Adaspeech [12] have further evolved TTS capabilities and improved TTS speech quality. Deepvoice2 [13] introduced the use of speaker verification models [14]–[16] to achieve multi-speaker TTS [17]–[21].

Child-TTS (CTTS), or TTS synthesis for child speech is currently limited due to the scarcity of child voice datasets and the challenges associated with their creation. Collecting child speech data for TTS research is challenging. Most TTS datasets are created in studios with expensive equipment, tailored for adult voices. While the pitch for adults typically falls between 70 to 250 Hz, children's speech ranges from 200 to 500 Hz [22]. Additionally, child speech exhibits distinct characteristics from adult speech, such as a higher fundamental frequency and variable speaking rates compared to adults [23]–[26]. Moreover, children tend to have longer phoneme durations and different prosody features due to their smaller vocal tracts [27]–[29].

This research aims to harness the potential of state-of-the-art (SOTA) TTS methods such as Fastpitch [11] to construct a pipeline for synthesizing children's voices while minimizing data requirements. The primary objective is to demonstrate the pipeline's ability to reliably generate a variety of self-consistent, distinct children's voices. Fastpitch utilizes a pitch prediction and duration prediction module which captures pitch variations in speech and enables more precise control over the speaking rate. This study uses an existing multispeaker children's speech dataset [30], which was cleaned to make it more suitable for CTTS research [31]. Subsequently, Fastpitch was trained on the cleaned dataset to generate synthetic speech for multiple child speakers, serving as a proof of concept.

By incorporating Fastpitch into the synthesis pipeline, we can effectively capture the unique prosodic features and intonation patterns present in speech. Our objective is to further optimize this model for child speech to accommodate individual characteristics, such as gender and regional accents, to produce realistic synthetic child voices. By using this approach, we intend to overcome the limitations of traditional TTS systems that often fail to capture the naturalness and authenticity of child-like speech. Our hypothesis is centered on the idea that pretraining the TTS model on adult speech data and subsequently finetuning it with child speech data can facilitate the synthesis of artificial child speech.

As part of this research, we also release a small set of synthetic datasets generated from this research. Objective evaluations were conducted on the synthesized child voices, comparing them to real child voices in terms of various acoustic features and Mean Opinion Score (MOS). The evaluation encompassed factors such as 'Naturalness', 'Intelligibility', and 'Speaker Similarity. Furthermore, we compared this approach with our previously reported Tacotron 2 TTS pipeline for the child speech synthesis [32]. In this study, no subjective evaluation was conducted; however, it will be taken into consideration for future research.

The potential applications of this research are wide-ranging and impactful such as educational tools, audiobooks for children, language learning, interactive games and toys, virtual learning companions, and child-friendly voice assistants and chatbots to name a few. Such a pipeline would also enable the creation of large synthetic datasets, which could, in turn, enhance other areas of child speech research, such as speaker recognition and automatic speech recognition [33], [34].

## II. METHODOLOGY

Fastpitch is a fully parallel TTS model conditioned on fundamental frequency contours. By incorporating Fastpitch into the synthesis pipeline, we can effectively capture the unique prosodic features and intonation patterns present in child speech. We present a multispeaker framework for TTS using a transfer learning approach that uses Waveglow vocoder for audio synthesis. We also evaluate this methodology using different objective evaluation methods to provide the validity of this approach. Fastpitch is used in this work due to its various advantages such as faster inference speed, improved prosody control, enhanced naturalness, duration control, multilingual support, and simplified architecture as compared to previous TTS approaches.

### A. Datasets

In this section, we give an overview of the datasets used to finetune our pipeline and to implement some of our evaluation methods.

*1) TTS Datasets:*

These datasets are used for the TTS experiments for pretraining and finetuning the TTS model.

**LibriTTS [35]:** The LibriTTS corpus is an adult speech dataset that includes 585 hours of speech data sampled at a rate of 24kHz, obtained from a diverse set of 2,456 speakers. LibriTTS is widely used in research for training and evaluating text-to-speech systems.

**MyST [30]:** My Science Tutor (MyST) Corpus [36] is an American English child speech dataset from 1371 students containing over 393 hours of audio data out of which 197 hours are fully transcribed. We use the cleaned version of this dataset (derived from [31]), with 65 hours of speech divided into two subsets: 55 hours for training, called 'MyST_train' and 10 hours for testing, called 'MyST_test'. This 55 hours of training data is used for TTS training.

*2) Text Datasets:*

These datasets are used during inference as input text for the TTS model to generate data samples from the finetuned synthetic child voices.

**Harvard Sentences [36]:** Harvard sentences consist of 720 sentences that are carefully designed to be phonetically balanced. These sentences effectively encompass a wide range of phonemes.

**LJ Speech Sentences [37]:** This dataset contains 13,100 sentences extracted from the LJ Speech dataset.

### B. Multispeaker child TTS using Fastpitch

*1) Fastpitch (Acoustic Model) [11]:*

FastPitch is a streamlined TTS model with a simplified encoder-decoder architecture, designed for faster inference and improved prosody control. In the multispeaker FastPitch TTS model, the input text is encoded using an encoder module, which typically comprises stacked layers of convolutional neural networks (CNNs) or recurrent neural networks (RNNs). The encoder processes the linguistic features of the text, such as phonemes or graphemes, and generates intermediate representations. The duration predictor module takes the intermediate representations from the encoder and predicts the duration of each phoneme or character in the input text. This enables the model to capture and generate natural speech rhythm and timing. The pitch predictor module takes the intermediate representations and predicts the fundamental frequency (F0) contour, controlling the pitch variations in the synthesized speech. The architecture of Fastpitch is detailed in Figure 1.

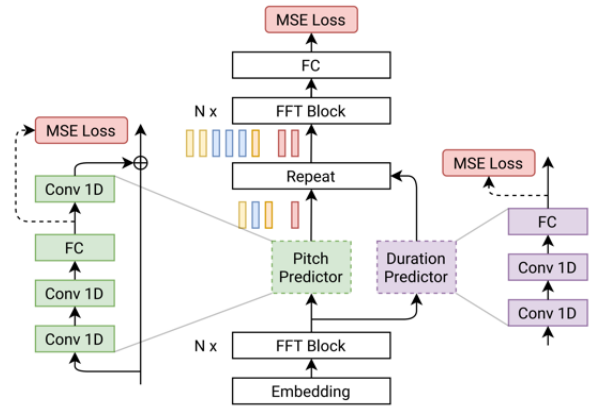

Fig. 1. Fastpitch Architecture [11].

We also condition the model on the speaker by adding a global speaker embedding [38] to the input tokens. The speaker embedding integration with the TTS framework [39]–[41] allows the model to capture the unique characteristics of different speakers. These embeddings encode speaker-related information in a vectorial representation for each speaker. During training, the model learns to associate speaker embeddings with the corresponding speakers, allowing it to generate speech that not only follows the desired linguistic content but also reflects the distinct vocal attributes of specific speakers. The primary loss function is the mean squared error (MSE) between the predicted mel-spectrogram and the target mel-spectrogram. Our work uses a newer version of Fastpitch, which is based on using the self-attention framework proposed in [38]. This enables the TTS model to learn speech-to-text alignment in parallel to TTS training instead of relying on an external aligner.

*2) Transfer Learning Pipeline:*

The proposed methodology involves pretraining the Fastpitch TTS model on a diverse dataset of adult speech, covering various age groups, linguistic backgrounds, and speech contexts. The LibriTTS dataset was used in this work. By finetuning the pretrained model on a smaller subset of the child speech dataset, such as MyST, will enable the model to learn the distinctive acoustic properties and pitch contours specific to child speech. Moreover, the model can be further optimized to accommodate individual characteristics, such as gender and regional accents, to synthesize more realistic CTTS voices.

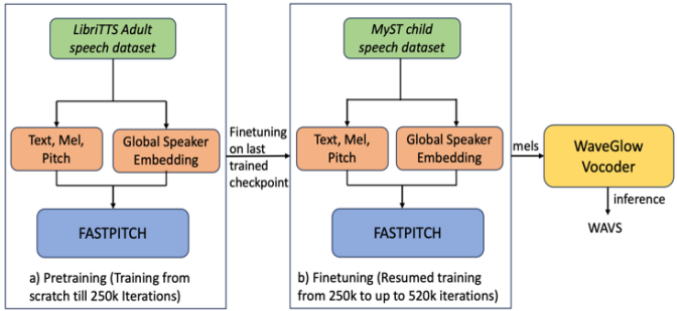

Fig. 2. Transfer learning pipeline: a) Pretraining: model being trained with LibriTTS dataset for up to 250k iterations. b) Finetuning: Resuming the acoustic model training with the MyST dataset from 250k iteration onwards up to 520k iteration.

The finetuning pipeline is kept consistent with our previous approach using Tacotron 2 [32] to allow for comparisons. Figure 2 describes the transfer learning pipeline. The model is first trained with the LibriTTS dataset (585 hours) for up to 250k iterations until a consistent low loss threshold is achieved, and the model starts to converge. After that, the model was finetuned for up to 520k additional steps using the MyST dataset (55 hours).

*3) Waveglow (Vocoder) [42]:*

WaveGlow is a SOTA vocoder model that generates high-quality and natural-sounding speech waveforms. It is based on a generative flow-based model architecture which models the distribution of speech waveforms. WaveGlow operates by taking a spectrogram representation of the speech as input and generating the corresponding waveform. The model employs an invertible neural network to transform the spectrogram into a latent space representation and then uses a series of invertible coupling layers to map this latent representation back into the waveform domain. Our WaveGlow model is trained on LibriTTS adult speech data which learns the complex relationship between spectrograms and waveforms. It was observed that Glow models [43]–[46] has popularly been used as a universal vocoder [45] and has been shown to work well with unseen speakers in multi-speaker models as well [47], [48]. Therefore, for the scope of this paper, WaveGlow (trained on LibriTTS) is used as a universal vocoder with synthetic child voices.

## III. EXPERIMENTS

### A. Training details

The implementation is obtained from Nvidia's FastPitch Github[1]. For our training and finetuning process, we utilized two A6000 40GB GPUs. We employed a learning rate of 0.1 and a weight decay factor of 1e-6, maintaining consistency with their original implementation [11]. Additionally, the remaining hyperparameters were retained as per the provided implementation details. To ensure a smooth training process, we incorporated a warmup training step with a factor of 2000.

### B. Experiments

*1) Initial Experiments:* These experiments involved using the LJ speech dataset for single-speaker finetuning. The model was first trained with LJ Speech and then finetuned with a single speaker from the MyST dataset. The output audio obtained was quite noisy. We also tried training the LJ speech single-speaker dataset and finetuning it with the complete MyST dataset (considering it as a single-speaker dataset). However, the results obtained didn't sound like child speech. Hence, finetuning on a single speaker was not explored further.

[1] https://github.com/NVIDIA/DeepLearningExamples/tree/master/PyTorch/SpeechSynthesis/FastPitch

*2) Main Experiments:* These experiments involve multispeaker TTS training. The model was first trained with the LibriTTS dataset. Figure 3 shows an example loss curve of the LibriTTS training.

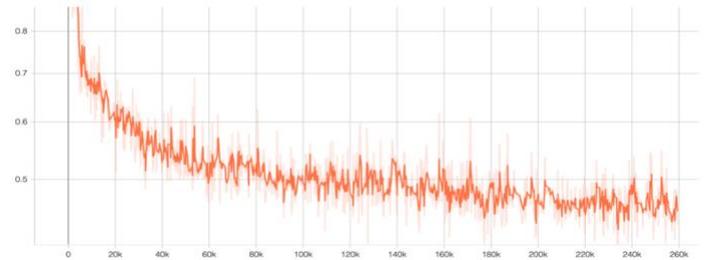

Fig. 3. LibriTTS pretraining curve (MSE loss vs. number of epochs)

It can be observed that for the first 2000 warmup steps, loss decreases gradually. After that loss decreases steadily until it reaches an average loss of 0.3 around 250k epoch. Since there was no improvement in loss function after that, it was decided to pause the training for further finetuning.

Further finetuning was performed from epoch 250k onwards on the MyST dataset. The loss increases until it starts to decrease around 260k epoch. From this point, there is a gradual decrease in loss until 520k steps. No significant improvement was observed in loss after this epoch. This was also verified by manually listening to generated audio files at an interval of every 50k epoch. After 550k epochs, the model exhibited signs of overfitting and began learning noise features in the MyST dataset, resulting in a decline in the quality of the synthesized audio.

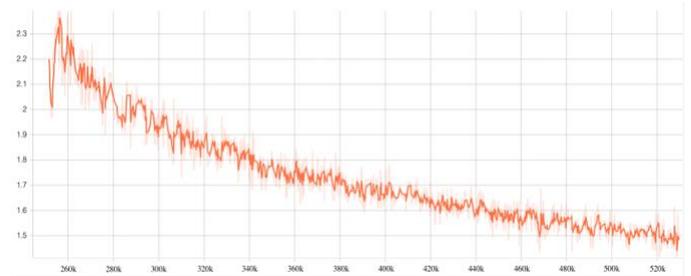

Fig. 4. MyST Finetuning curve (MSE loss vs. number of epochs).

### C. Synthetic Datasets

We have generated two sets of synthetic child speech datasets. The dataset demographic is detailed in Table III. The dataset is made available through our GitHub[2]. Since the dataset was generated at a 22Khz sampling rate, FFmpeg was used to convert the data into a 16khz sampling rate for objective evaluation. The dataset is made available in both sampling rates. The dataset details are available below:

[2] https://github.com/C3Imaging/child_tts_fastpitch/

TABLE I. SYNTHETIC DATASET DEMOGRAPHICS

| Dataset | Speakers | Hours | Utterances | data/speaker |
|---|---|---|---|---|
| CS_HS | 40 | 29.02 | 28,800 | 43.53 minutes |
| CS_LJ | 2 | 47.61 | 26,200 | 23.8 hours |

*1) CS_HS* – This dataset used Harvard Sentences as a text reference to generate the synthetic child speech dataset. We selected the 40 speakers with the most amount of data in hours. from the LibriTTS dataset which was used to generate 40 child speakers. See Table 1 for more details.

*2) CS_LJ – This dataset used* LJ Speech transcripts as a text reference to generate the synthetic child speech dataset. We selected one male and one female speaker from the LibriTTS dataset which contained the most amount of training. These speakers were subjected to generate the child's speech. See Table 1 for more details.

## IV. RESULTS AND EVALUATION

Our experimental findings demonstrate the successful synthesis of child voices using our proposed methodology. To assess the validity of the generated speech, we conducted objective evaluations, specifically focusing on the aspects of Naturalness, Intelligibility, and Speaker similarity. Furthermore, we conducted a comparative analysis with our previous research, which involves synthesizing child speech using the Tacotron 2 model. For the evaluation process, we randomly selected 120 utterances from the original MyST dataset, Tacotron-based synthetic dataset, and Fastpitch-generated synthetic utterances (from III.B). This allowed us to systematically compare the quality of speech generated by both the Tacotron 2 [32] and Fastpitch models within the context of child speech synthesis.

### A. Objective Naturalness Evaluation using the pretrained MOSNet [49]

TABLE II. MOSNET OUTPUT FOR 120 SAMPLES WITH 95% CONFIDENCE INTERVAL

| Dataset | MOS |
|---|---|
| Adult speech (Librispeech test_clean) | 3.78 ± 0.07 |
| Original Child Speech [MyST] | 2.91 ± 0.07 |
| Tacotron 2 based synthetic child speech [32] | 2.60 ± 0.06 |
| **Fastpitch based synthetic child speech [Our work]** | **3.10 ± 0.12** |

Table 1 provides the Mean Opinion Scores (MOS) for 120 different speech samples using the pretrained MOSNet model [49]. MOSNet, trained on adult speech, exhibits a high correlation with human MOS ratings. However, its generalization to child speech is doubtful. Therefore, we only use MOSNet in this study to explore the correlation between reference child audio and synthetic child audio. It acts as a measure to validate the 'Naturalness' of the speech The original child speech from the MyST dataset received an average MOS of 2.91 ± 0.07, indicating moderate acceptability. The Fastpitch-generated child speech indicates higher quality than both the original speech and Tacotron2. These results suggest that the Fastpitch model, as implemented in our research, produces a strong correlation between synthetic child speech and real child speech.

### B. Objective Intelligibility Evaluation using a pretrained wav2vec2 ASR System [50]

TABLE III. WER ON 120 RANDOMLY SELECTED UTTERANCES FROM ADULT SPEECH, REAL CHILD SPEECH, AND SYNTHETIC CHILD SPEECH USING THE WAV2VEC2 BASE ASR MODEL

| Dataset | WER |
|---|---|
| Adult Speech (Librispeech test_clean) | 3.43 |
| Original Child Speech [MyST] | 15.27 |
| Tacotron 2 based synthetic child speech [32] | 25.63 |
| **Fastpitch based synthetic child speech [Our work]** | **17.61** |

In this study, we employed the wav2vec2 base model [3], which was finetuned with 960 hours of the Librispeech dataset, to evaluate the 'Intelligibility' of the generated child speech. Since wav2vec2 is a SOTA ASR model, it was intended to use this as a validity metric for the synthetic speech. Additionally, we conducted a comparative analysis with our previous approach utilizing Tacotron 2 [32]. Table II provides WER for different speech datasets. The adult speech dataset achieved a strong WER of 3.43, considering the model's training on adult speech data. Our Fastpitch-based approach achieved a WER of 17.61, closely resembling the WER of the original child speech from the MyST dataset. Moreover, it surpassed the WER of the Tacotron 2 generated child speech, indicating improved performance over the synthetic child speech.

### C. Speaker similarity verification using a pretrained speaker verification system [15]

Speaker similarity between a synthesized speech and a real speech can be calculated using a speaker verification system [15]. The pretrained speaker encoder from Resemblyzer[4] was used to extract and visualize the speaker embeddings. This tool uses cosine distance to calculate the similarity between the two embeddings.

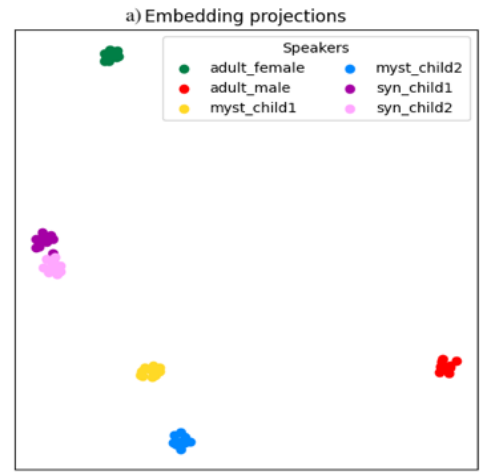

---

[3] https://github.com/facebookresearch/fairseq/tree/main/examples/wav2vec

[4] https://github.com/resemble-ai/Resemblyzer

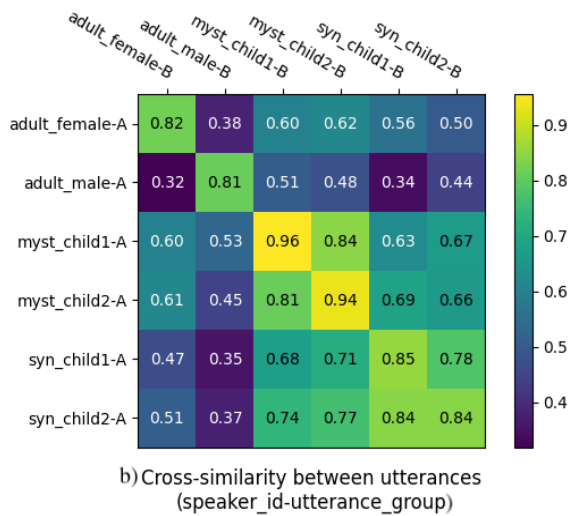

Fig. 5. a) Projections of embeddings between different real and synthetic child speech in comparison to adult speakers. b) Cross-similarity between 10 speakers in Set A and Set B.

For this evaluation, 6 speakers were randomly selected: 2 from LibriTTS [one male and one female], 2 from the MyST dataset, and 2 from the synthetically generated CS_HS dataset. We selected 10 utterances for each speaker in a random order. A visualization of this similarity in a 2D projection can be seen in Figure 5a. It can be observed that most of the child speakers (both real and synthetic) are very close in a cluster compared to adult male and female speakers.

To further demonstrate the similarity between real child speech and synthetic child speech, cosine similarity was used to calculate the cross-similarity between each speaker. All 6 speakers were divided into two sets, A and B. Speaker embeddings are extracted for each of the utterances for each of the sets and averaged together for each speaker. This gave us 6 unique speaker embeddings in sets A and B for each of the 6 speakers. Cosine similarity is finally used to measure the similarity between sets A and B. Figure 5b shows the plot for the cross similarity between 6 speakers. The similarity for most of the child and adult speech is between 0.34-0.53 whereas the similarity for synthetic child speech and real child speech is between 0.63-0.98. The average similarity between synthetic and real child voices is 77%. Hence, we can conclude that our synthetically generated child speech is quite close to real child speech in terms of speaker similarity.

## V. CONCLUSION AND FUTURE WORK

This paper presents a pipeline for synthesizing child speech in scenarios with limited training data. The proposed approach involves cleaning an existing child speech dataset to create a small, curated dataset suitable for TTS training. A transfer learning technique is employed, utilizing pretraining on adult speech data and finetuning on child speech data. Objective evaluations using MOSNet demonstrate a strong correlation between real and synthesized child voices. Using a pretrained adult speech wav2vec2 ASR model, the WER for synthetic child voices was measured at 17.61, compared to a WER of 15.27 for real child voices. Speaker similarity evaluation using a pretrained speaker encoder yields an average cosine similarity of 77% between synthetic speech and the original speakers. Synthetic child speech samples are available on the project's GitHub. We also release two small synthetic child speech datasets generated from this work. Multi-speaker TTS proves to be a valuable approach for child speech synthesis, even with limited training data.

For Future work, we aim to perform a subjective evaluation (as proposed in [32]) on the released dataset for better clarity over the 'Naturalness', 'Intelligibility', and 'Speaker Similarity' of the generated child speech. Furthermore, it is also intended to investigate the use of synthetically generated child speech to enhance other areas of child speech research, such as ASR and speaker recognition.

ACKNOWLEDGMENT

The authors would like to acknowledge experts from Xperi: Gabriel Costache, Zoran Fejzo, Francisco Salgado, and George Sterpu for providing their expertise and feedback throughout. The authors would also like to thank Adriana Stan and Horia Cucu from the University Politehnica of Bucharest, for providing her expertise on TTS/ASR experiments.